\documentclass[12pt]{article}
\newif\iffigs\figstrue

\usepackage{epsfig,latexsym}
\usepackage{amsmath}
\usepackage{color}
\usepackage{graphicx}
\usepackage{verbatim}
\usepackage{mathrsfs}
\usepackage{amssymb}

\DeclareMathAlphabet{\mathpzc}{OT1}{pzc}{m}{it}

 \csname
@addtoreset\endcsname{equation}{section}

\def\gz0{\gamma^{0}}

\def\bec{\begin{center}}
\def\ec{\end{center}}

\def\12{\frac{1}{2}}

\def\DH{\rm I\kern-1.5pt\rm H\kern-1.5pt\rm I}

\def\DR{\rm I\kern-1.45pt\rm R}
\def\DC{\kern2pt {\hbox{\sqi I}}\kern-4.2pt\rm C}

\newcommand{\beq}{\begin{equation}}
\newcommand{\eeq}{\end{equation}}
\newcommand{\bea}{\begin{eqnarray}}
\newcommand{\eea}{\end{eqnarray}}
\newcommand{\bi}{\begin{itemize}}
\newcommand{\ei}{\end{itemize}}

\usepackage{slashed}

\usepackage{float}

\usepackage{caption}

\usepackage{epsf}
\usepackage{amsmath}

\definecolor{alizarin}{rgb}{0.82, 0.1, 0.26}

\usepackage{multirow}

\usepackage{epsfig}
\usepackage{cite}
\usepackage{color,colordvi}

\newcounter{hran}

\makeatletter
\renewcommand\section{\@startsection {section}{1}{\z@}%
                               {-3.5ex \@plus -1ex \@minus -.2ex}%
                               {2.3ex \@plus.2ex}%
                               {\normalfont\large\bfseries}}
\makeatother

\vspace{0.5cm}

\begin{document}
\thispagestyle{empty}

\vspace{15pt}

\begin{center}


{\Large\sc An Update on Brane Supersymmetry Breaking }\\


\vspace{50pt}
{\sc J.~Mourad${}^{\; a}$  \ and \ A.~Sagnotti${}^{\; b}$}\\[15pt]

{${}^a$\sl\small APC, UMR 7164-CNRS, Universit\'e Paris Diderot -- Paris 7 \\
10 rue Alice Domon et L\'eonie Duquet \\75205 Paris Cedex 13 \ FRANCE
\\ }e-mail: {\small \it
mourad@apc.univ-paris7.fr}\vspace{8pt}

{${}^b$\sl\small
Scuola Normale Superiore and INFN\\
Piazza dei Cavalieri, 7\\ 56126 Pisa \ ITALY \\
e-mail: {\small \it sagnotti@sns.it}}\vspace{10pt}

\vspace{16pt}

\vspace{25pt} {\sc\large Abstract}
\end{center}

\noindent
``Brane supersymmetry breaking'' is a peculiar phenomenon that can occur in perturbative orientifold vacua. It results from the simultaneous presence, in the vacuum, of non--mutually BPS sets of BPS branes and orientifolds, which leave behind a net tension and thus a runaway potential, but \emph{no} tachyons. In the simplest ten--dimensional realization, the low--lying modes combine the closed sector of type--$I$ supergravity with an open sector including $USp(32)$ gauge bosons, fermions in the antisymmetric 495 and an additional singlet playing the role of a goldstino.  We review some properties of this system and of other non--tachyonic models in ten dimensions with broken supersymmetry, and we illustrate some puzzles that their very existence raises, together with some applications that they have stimulated.

\vskip 45pt

\begin{center}
{\sl Based in part on the lectures presented by A.S. at the \\ ``International School of Subnuclear Physics'', 55th Course, Erice, June 14 -- 23 \ 2017, \\ at TFI-2017, Parma, September 11 -- 13 \ 2017, \\ and at LACES 2017, Firenze, November 27 -- December 1, 2017}
\end{center}
\vskip 45pt

\begin{center}
{\sl Dedicated to the memory of Yassen S.~Stanev}
\end{center}

\vfill
\noindent

\baselineskip=15pt

\newpage

\setcounter{page}{2}
\setcounter{equation}{0}
\section{Introduction} \label{sec:intro}

There are only a few ten--dimensional string models~\cite{stringtheory} without tachyonic excitations, and they fall naturally into three classes, depending on the number of supersymmetries. The first is widely familiar: it comprises the $IIA$ and $IIB$ models, whose low--lying modes belong to the unique type--$IIA$ $(1,1)$ and type--$IIB$ $(2,0)$ supersymmetry multiplets, and whose low--energy interactions are encoded in the corresponding versions of ten--dimensional supergravity~\cite{supergravity}. Another class, which is equally familiar, comprises the two heterotic $HE$ and $HO$ models~\cite{heterotic}, with gauge groups $E_8 \times E_8$ and $SO(32)$, and the type--$I$ $SO(32)$ (open and closed) superstring~\cite{greenschwarz}. In these three cases, the low--lying modes fill the two available types of $(1,0)$ supersymmetry multiplets, while the low--energy interactions are encoded in minimal supergravity coupled {\it \`a} la Green--Schwarz~\cite{greenschwarz} to ten--dimensional supersymmetric Yang--Mills theory. However, the three models are intrinsically different, since the $SO(32)$ (open and closed) superstring provides the simplest example of an orientifold construction~\cite{orientifolds}, which links it to the type--$IIB$ model. A third class of non--tachyonic models exists, however, which is the focus proper of this review. Supersymmetry is either absent in their tachyon--free spectra or it is strikingly present in a non--linear phase, without an order parameter capable of recovering it. The first model of this type is the $SO(16) \times SO(16)$ heterotic string of~\cite{so16xso16}: as we shall review shortly, its chiral spectrum obtains from the supersymmetric $HE$ model via an orbifold construction, in the spirit of~\cite{orbifolds}. The second model descends via an orientifold construction~\cite{u32} from a tachyonic and purely bosonic model of oriented closed strings~\cite{type0}: its low--lying spectrum is again chiral, and hosts a $U(32)$ gauge group~\footnote{The unbroken gauge symmetry is actually $SU(32)$, as discussed in the second paper of~\cite{u32}.}. Finally, the third model stands out as potentially the most puzzling of all: its partition function is almost identical to the one of the type--$I$ superstring, and yet supersymmetry is broken in it at the string scale. More precisely, supersymmetry is non--linearly realized, consistently with the presence of a gravitino in its low--lying chiral spectrum, which also hosts a $USp(32)$ gauge group~\cite{sugimoto}. The $USp(32)$ string model also provides the simplest instance of ``brane supersymmetry breaking''~\cite{bsb}: in Polchinski's space--time picture~\cite{polchinski}, its vacuum hosts non--mutually BPS combinations of BPS branes and orientifolds, and no tachyons arise due to the non--dynamical nature of the latter. The mere existence of this model is thus a puzzle and a challenge by itself, while its dynamics raises questions and hopes that make it a paradigm for further searches aimed at reaching beyond the celebrated hexagon of $M$--theory dualities~\cite{10Ddualities}.

Quantum corrections in models of oriented closed strings, or already the tree--level interactions among non--mutually BPS string solitons present in the vacuum, manifest themselves via the emergence of a runaway dilaton potential in the low--energy effective field theory. One is typically left with a net attractive force which would tend to drive the vacuum toward collapse. Difficulties are encountered ubiquitously in searches for static vacua, while cosmological solutions may appear more natural in this context, and have indeed some striking features. More precisely, obtaining solutions which are stable and have bounded quantum and world--sheet string corrections
is still a challenge. In time--dependent backgrounds, as we shall see, one can attain bounded quantum corrections, while in backgrounds with fluxes both types of corrections can be small. In the former case world--sheet string corrections remain a possible issue, while in the latter stability is not guaranteed.

This review is an attempt to summarize briefly the main properties of these systems. Our aim is to highlight some surprising indications and stimuli that models with a high--scale string breaking of supersymmetry have already provided, despite the difficulties we have alluded to, while stressing some of the many open questions that remain about their dynamics. In Section \ref{sec:10D} we review the construction of the partition functions for tachyon--free ten--dimensional models. In Section \ref{sec:classical_sols} we review the main properties of some classical solutions driven by the runaway potentials that have been analyzed in detail, and in Section \ref{sec:springoffs} we summarize some important steps that led to the formulation of four--dimensional toy models with a high scale of supersymmetry breaking via constrained superfields, and we conclude with what can be regarded as the simplest four--dimensional toy model for brane supersymmetry breaking.

\section{Non--tachyonic ten--dimensional strings} \label{sec:10D}
Characterizing string spectra is a particularly important step, since perturbative interactions follow directly via insertions on Riemann surfaces of local operators corresponding to their states. For oriented closed strings in ten dimensions, there are only a few available options, which can be identified referring to the torus amplitude and relying on two main tools. The first is the Neveu--Schwarz--Ramond light--cone formulation~\cite{NSR}, while the second is the Gliozzi--Scherk--Olive (GSO) projection~\cite{GSO}. The former provides the (Bose and Fermi) two--dimensional oscillator modes which build string states, while the latter selects the subsets of excitations that are effectively consistent with the two key requirements of \emph{modular invariance} and \emph{spin--statistics}.

In approaching the GSO projection, it is very convenient to work with $SO(8)$ level--one characters, which encode the independent sectors of the spectrum with definite spin--statistics properties both in space time and on the string world sheet. These characters are a special case of more general (level--one) $SO(2n)$ characters that is selected by the manifest transverse $SO(8)$ rotation symmetry available in ten--dimensional Minkowski space. One can build indeed four distinct sectors of states acting on the vacua of the antiperiodic (Neveu--Schwarz) or periodic (Ramond) sectors, for both left--moving and right--moving oscillators when they are available. This situation would extend to all $SO(2n)$ groups, where the first two characters, $O_{2n}$ and $V_{2n}$, would count states built with even or odd numbers of Neveu--Schwarz oscillators acting on the corresponding vacuum. On the other hand the last two characters, $S_{2n}$ and $C_{2n}$, would count states built acting on the Ramond vacuum with corresponding oscillators while also enforcing opposite choices of alternating chiral projections at all levels. In this fashion, say, $S_{2n}$ would involve left chiral projections at all odd levels and right chiral projections at all even ones, while these projections would all be reversed in $C_{2n}$. Massive light--cone spectra would then combine nonetheless, as expected for Lorentz--invariant spectra, into non--chiral massive ten--dimensional multiplets. These characters,
\begin{eqnarray}
& O_{2n} \ = \ \frac{\theta^n\left[\substack{0\\0}\right]\left(0|\tau\right)\, + \ \theta^n\left[\substack{0\\{1/2}}\right]\left(0|\tau\right)}{2\,\eta^n(\tau)}\ ,  \qquad S_{2n} \ = \ \frac{\theta^n\left[\substack{{1/2}\\0}\right]\left(0|\tau\right)\, + \ i^{-n}\, \theta^n\left[\substack{{1/2}\\{1/2}}\right]\left(0|\tau\right)}{2\,\eta^n(\tau)} \, , \nonumber\\
& V_{2n} \ = \ \frac{\theta^n\left[\substack{0\\0}\right]\left(0|\tau\right)\, - \ \theta^n\left[\substack{0\\{1/2}}\right]\left(0|\tau\right)}{2\,\eta^n(\tau)}\ ,  \qquad C_{2n} \ = \ \frac{\theta^n\left[\substack{{1/2}\\0}\right]\left(0|\tau\right)\,- \, i^{-n}\, \theta^n\left[\substack{{1/2}\\{1/2}}\right]\left(0|\tau\right)}{2\,\eta^n(\tau)} \, , \nonumber\\
& \eta(\tau) \ = \ q^\frac{1}{24}\ \prod_{n=1}^\infty (1 \ - \ q^n ) \, , \qquad q \,=\, e^{2\pi i \tau}\,,\nonumber \\  & \theta\left[\substack{\alpha\\\beta}\right]\left(z|\tau\right) \ = \ \sum_{n \in Z} \ q^{\,\frac{1}{2}\, (n+\alpha)^2} \, e^{i 2 \pi(n+\alpha)(z-\beta)} \ , \label{characters}
\end{eqnarray}
are combinations of Jacobi $\theta$--functions with characteristics~\cite{theta_review} and the Dedekind $\eta$ function, which is also needed to encode the contributions of bosonic oscillators. The torus amplitude can be defined working on the complex plane with the two identifications $z \sim z+1$ and $z \sim z+\tau$. The corresponding modular transformations act on $\tau$ via the fractional linear transformations
\beq
\tau \ \to \ \frac{a\, \tau \ + \ b}{d\, \tau \ + \ d} \qquad (ad \,-\, bc \,=\,1) \ , \label{sl2z}
\eeq
and can be built out of two generators $(S : \tau \to \,- \ \frac{1}{\tau}, \ T: \tau \to \tau + 1)$. $S$ and $T$ act on the four characters via the two matrices
\beq
 S \ = \ \frac{1}{2} \left( \begin{array}{rrrr} 1 & 1 & 1 & 1 \\
1 & 1 & -1 & -1 \\
1 & -1 & i^{-n} & -i^{-n} \\
1 & -1 & -i^{-n} & i^{-n}  \end{array} \right)\, ,  \quad T \ = \ e^{\,-\,\frac{i n \pi}{12}} \ \left( \begin{array}{rrrr} 1 & 0 & 0 & 0 \\
0 & -1 & 0 & 0 \\
0 & 0 & e^{\frac{i n \pi}{4}} & 0 \\
0 & 0 & 0 & e^{\frac{i n \pi}{4}}  \end{array} \right)\,  .  \label{modular}
\eeq
and on the Dedekind function as $\eta\left(\,-\,1/\tau \right) \ = \ (-i\,\tau)^\frac{1}{2} \ \eta (\tau)$ and $\eta(\tau+1) \ = \ e^{\,\frac{i \pi}{12}}\ \eta(\tau)$.

It is important to appreciate that the definitions in eq.~\eqref{characters} rest on a key step, whereby \emph{physically different} but \emph{numerically identical} contributions are distinguished. In two--dimensional Conformal Field Theory this step is usually called ``resolution of the ambiguity''. In the present examples the ``odd spin structure'' contribution $\theta\left[\substack{{1/2}\\{1/2}}\right]\left(0|\tau\right)$ vanishes, consistently with the fact that it involves a counterpart of the four--dimensional chirality matrix $\gamma_5$. Still, one can consistently distinguish the $S_{2n}$ and $C_{2n}$ sectors, and this choice has the additional virtue of bringing the matrix $S$ into the symmetric and unitary form of eq.~\eqref{modular}. In more complicated examples of two--dimensional Conformal Field Theory, this procedure can actually reveal the presence of different sectors, which here is evident for physical reasons. We can now review how modular invariance, which is tantamount to invariance of the torus partition function under the two transformations in eqs.~\eqref{characters}, when combined with properly opposite signs for the contributions of space--time Bose and Fermi modes, can neatly select consistent spectra of oriented closed strings.

In the ten--dimensional Minkowski background that sets the stage for this construction, supersymmetry demands that equal numbers of Bose and Fermi excitations be present at every mass level. In this formalism, one can verify when this condition holds enforcing on partition functions Jacobi's \emph{aequatio}, which takes the form
\beq
V_8 \ = \ S_8 \ = \ C_8   \label{aequatio}
\eeq
in our notation. The partition functions have indeed a dual role: they encode neatly and efficiently perturbative string spectra in the terms satisfying ``level--matching'', \emph{i.e.} involving equal powers of $q$ and ${\bar q}$ of their Taylor expansions, but they also compute the vacuum energy that these states induce at the one loop (torus) level, once eq.~\eqref{aequatio} is used. When the total contribution does not vanish, which is typically the case with broken supersymmetry up to a few known exceptions~\cite{vanishing_nonsusy}, the remainder signals a back--reaction of the string spectrum on space time~\footnote{There are recent examples of closed--string models where a Bose--Fermi degeneracy at the massless level leads to an exponential suppression of the vacuum energy even in the presence of broken supersymmetry~\cite{exp_sup}.}. Unfortunately, there is no convenient way to account for these effects, as of today, for the string as a whole, although this key issue was identified long ago~\cite{vacuum_redefinitions}. As a result, strictly speaking the formalism is well under control only in supersymmetric constructions. If the energy scales at stake for the deformed backgrounds were to lie far below the scale of string excitations, the low--energy effective field theory could be expected to provide a reliable tool to characterize them, but unfortunately this nice state of affairs is more an exception than the rule, and in particular it does not present itself in brane supersymmetry breaking, which is the phenomenon of primary interest for this review.

Let us begin by describing the supersymmetric models, which do not bring along this type of subtlety, before getting to broken supersymmetry. Our first examples are thus the torus partition functions of the $IIA$ and $IIB$ superstrings, which read
\beq
{\cal T}_{IIA} \ = \ \int_{\cal F} \ \frac{d^2 \tau}{(Im \tau)^2} \ \frac{(V_8 \,-\,S_8)(\bar{V}_8 \,-\,\bar{C}_8)}{(Im \tau)^4 \ \eta^8 \, \bar{\eta}^8} \, , \quad {\cal T}_{IIB} \ = \ \int_{\cal F} \ \frac{d^2 \tau}{(Im \tau)^2} \ \frac{\left|V_8 \,-\,S_8\right|^2}{(Im \tau)^4 \ \eta^8 \, \bar{\eta}^8} \ \ . \label{IIA_IIB}
\eeq
The reader may verify by inspection their invariance under the $S$ and $T$ transformations of eq.~\eqref{modular}, while also noticing the subscript in the two integrals, which is well familiar to string experts. It identifies a fundamental region of the modular group $SL(2,Z)$, which acts on $\tau$  as in eq.~\eqref{sl2z}. A conventional choice for ${\cal F}$ is the region of the $\tau$ plane between the two vertical lines at $Re(\tau)=\pm 1/2$ and above the unit circle centered at the origin, where modular invariance identifies the two vertical lines and two halves of the circular portion. Notice that the restriction to $|\tau| > 1$ testifies the absence of an ultraviolet problem in these models, while the presence in ${\cal F}$ of arbitrarily high values of $Im(\tau)$ leaves around subtle infrared effects.

Our compact notation has the virtue of highlighting the crucial difference between the two partition functions in eq.~\eqref{IIA_IIB}: the $IIB$ partition function is \emph{symmetric} under complex conjugation, which interchanges left--moving and right--moving modes, while its low--lying excitations are \emph{chirally asymmetric} in space--time. The opposite is true for the $IIA$ partition function,
which is \emph{asymmetric} on the world sheet while its low--lying excitations are \emph{chirally symmetric} in space--time. Keeping in mind only a few facts, the low--lying spectra can be read directly from eqs.~\eqref{IIA_IIB}. To begin with, one should note that the product $V_8 \, \bar{V}_8$ brings along the massless modes of a graviton, a dilaton and an antisymmetric two--form. The last arises from the antisymmetric combination of a pair of $NS$ oscillators, $\psi_{-\frac{1}{2}}^i\, {\overline \psi}_{-\frac{1}{2}}^{\,j}$, while the first two arise, respectively, from their traceless symmetric combination and the trace. In a similar fashion, the proper interpretation of $V_8 \, \bar{S}_8$ or of its conjugate also entails a split into a pair of irreducible contributions: the modes of a left gravitino lie in its $\gamma$--traceless portion, while a right spinor emerges from the $\gamma$ trace. A similar decomposition would exhibit the content of $V_8 \, \bar{C}_8$ or of its conjugate, while for the $RR$ states one has to keep in mind that $S_8 \, \bar{S}_8$ builds states whose light--cone potentials fold into even--order antisymmetric powers of $\gamma$--matrices. These modes correspond to a zero--form $\phi^\prime$, a two--form $B_{\mu\nu}^\prime$ and a four--form $D_{\mu\nu\rho\sigma}$, where the last gauge potential is subject to the peculiar constraint of having a self--dual field strength. Finally, $S_8 \, \bar{C}_8$ bring along a one--form potential $A_\mu$ and a three--form potential $C_{\mu\nu\rho}$~\footnote{One can also see that $O_8 \, \bar{O}_8$ starts with a tachyon, $C_8 \, \bar{C}_8$  starts with a set of fields that differ from those entering $S_8 \, \bar{S}_8$ because the field strength of the five-form is antiself--dual, while $C_8 \, \bar{S}_8$ starts as $S_8 \, \bar{C}_8$.}. All in all, the massless $IIA$ and $IIB$ spectra combine into the modes associated to the two sets of fields
\begin{eqnarray}&& { IIA: \ ( e_\mu^a,B_{\mu\nu},\phi,\psi_{\mu\,L},\psi_{\mu\,R},\psi_L,\psi_R, A_\mu,C_{\mu\nu\rho} )} \ , \nonumber \\
&& {IIB: \ ( e_\mu^a,B_{\mu\nu}^{1,2},\phi^{1,2},\psi_{\mu\,L}^{1,2},\psi_R^{1,2}, D_{\mu\nu\rho\sigma}^+ )}\ . \label{IIAB_spectra}
\end{eqnarray}
These massless modes are accompanied by infinite towers of excitations with squared masses proportional to the string scale $1/\alpha^\prime$, which is the only scale (implicitly) present in the models described in this section. For brevity, we have set indeed $\alpha^\prime=1$, and we shall continue to do it in most of the ensuing discussion. Starting from the light--cone partition functions of eq.~\eqref{IIA_IIB}, one can in principle reconstruct the whole massive spectrum of type--$IIA$ and type--$IIB$ strings.

The right--moving content of heterotic strings affords equivalent formulations via chiral internal bosons or fermions. The two are equivalent up to the Bose--Fermi two--dimensional mapping, which translates, for partition functions, into the replacement of the series expansions of eq.~\eqref{characters} for the $\theta$ functions with corresponding product decompositions. These models take again, in our notation, a simple form:
\beq
 {\cal T}_{HE} \, = \, \int_{\cal F} \ \frac{d^2 \tau}{(Im \tau)^2} \ \frac{(V_8 - S_8)(\bar{O}_{16} + \bar{S}_{16})^2}{(Im \tau)^4 \ \eta^8 \, \bar{\eta}^8} \ , \ {\cal T}_{HO} \, = \, \int_{\cal F} \ \frac{d^2 \tau}{(Im \tau^2)} \ \frac{(V_{8} - S_{8})(\bar{O}_{32} + \bar{S}_{32})}{(Im \tau)^4 \ \eta^8 \, \bar{\eta}^8} \ . \label{heterotics}
 \eeq
Analyzing their spectra requires however that the level--matching condition be carefully enforced: physical states are only associated to contributions whose mass levels computed from left or right modes coincide. Here the left--moving sector starts at the massless level, while the right--moving one starts at level -1. In general, $O_{2n}$ adds zero (plus integers) to this counting, since it contains the $NS$ vacuum, while $V_{2n}$ adds 1/2 (plus integers), since it includes at least one $NS$ oscillator. In a similar fashion, $S_{2n}$ and $C_{2n}$ add $n/8$ units (plus integers), a shift that reflects the overall power of $q$ carried by $\theta\left[\substack{{1/2}\\{0}}\right]\left(0|\tau\right)$. In the $HE$ model there are four internal fermionic sectors that can match the left sector at zero mass. To begin with, there are the lowest mass levels of $\bar{O}_{16}\,\bar{S}_{16}$ and $\bar{S}_{16}\,\bar{O}_{16}$, which build the $(1,128)$ and $(128,1)$ of $SO(16) \times SO(16)$, but there are also level--matching states at zero mass level in $\bar{O}_{16}\,\bar{O}_{16})$, if one acts on their tachyonic ground states with an (antisymmetric) pair of $NS$ oscillators $\psi_{-1/2}^I\,\psi_{-1/2}^J$ from the first factor and none from the second, or with an (antisymmetric) pair from the second factor and none from the first. These build the adjoint of $SO(16) \times SO(16)$, which together with the other states that we have identified reconstructs the adjoint of $E_8 \times E_8$. The $HO$ spectrum can now be identified by inspection, since the lowest mass level of ${\bar{S}}_{32}$ is lifted by two units, and by level matching cannot contribute at the massless level. One is thus left with the $\bar{O}_{32}$ sector, where taking once more (antisymmetric) pairs $\psi_{-1/2}^I\,\psi_{-1/2}^J$ of $NS$ oscillators build the adjoint of $SO(32)$ to be carried by the supersymmetric Yang--Mills multiplet. Moreover, in both cases combining the states built by a single right--moving bosonic oscillator with the left--moving sector builds the $(1,0)$ supergravity multiplet. All in all, one can thus associate the massless spectra of both models to the two sets of fields
 \begin{eqnarray}&& { HE: \ ( e_\mu^a,B_{\mu\nu},\phi,\psi_{\mu\,L},\psi_R) \oplus [ E_8 \times E_8 \ (A_\mu,\lambda_L)]}\ , \nonumber \\
&& {HO: \ ( e_\mu^a,B_{\mu\nu},\phi,\psi_{\mu\,L},\psi_R) \oplus [ SO(32) \ (A_\mu,\lambda_L)]} \ . \label{heterotic_spectra}
\end{eqnarray}
These correspond, as we have anticipated, to the $(1,0)$ supergravity and to the supersymmetric Yang--Mills theory in ten dimensions, with an $E_8 \times E_8$ gauge group for the $HE$ string and an $SO(32)$ gauge group for the $HO$ string.

The models that we have just described exhaust the available options for supersymmetric oriented closed strings. Another supersymmetric model, of a different type, exists however: it is the type--$I$ $SO(32)$ superstring, which is not really independent since it is an \emph{open descendant} or \emph{orientifold} of type $IIB$~\cite{orientifolds}. We shall come to it shortly, since the simplest realization of our main target, brane supersymmetry breaking, is apparently a mild variant of it. Before getting there, however, let us describe another heterotic model, which is not supersymmetric and yet is free of tachyons. This is the $SO(16) \times SO(16)$ heterotic string~\cite{so16xso16}, whose partition function reads
\bea
{\cal T}_{SO(16)\times SO(16)} \!\!\!&=&\!\!\! \int_{\cal F} \ \frac{d^2 \tau}{(Im \tau)^2} \ \frac{1}{(Im \tau)^4 \ \eta^8 \, \bar{\eta}^8} \  \left[O_8 (\bar{V}_{16}\,\bar{C}_{16}+\bar{C}_{16}\,\bar{V}_{16}) +
V_8 (\bar{O}_{16}\,\bar{O}_{16}\,+\,\bar{S}_{16}\,\bar{S}_{16}) \right. \nonumber \\ &-&\left.
S_8 (\bar{O}_{16}\,\bar{S}_{16}\,+\,\bar{S}_{16}\,\bar{O}_{16}) -
C_8 (\bar{V}_{16}\,\bar{V}_{16}\,+\,\bar{C}_{16}\,\bar{C}_{16})\right] \ . \label{so16so16}
\eea
The preceding discussion should help the reader to identify its low--lying field content,
\begin{eqnarray}&& { H_{16 \times 16}: \ ( e_\mu^a, B_{\mu\nu}, \phi) \oplus A_\mu^{(120,1)\oplus(1,120)} \oplus \psi_L^{(128,1)\oplus(1,128)} \oplus \psi_R^{(16,16)} } \ . \label{so16so16_spectrum}
\end{eqnarray}
Notice that these massless fields do not include a gravitino, as pertains to spectra not connected to supersymmetry, but interestingly no tachyons are present, since the tachyonic ground state of $O_8$ in the first group of contributions to eq.~\eqref{so16so16} does not satisfy level matching with its right--moving partners, whose spectrum begins 3/2 units above the ground state.

Here we encounter a first manifestation of a ubiquitous problem with broken supersymmetry: the partition function in eq.~\eqref{so16so16} does not vanish after enforcing in it eq.~\eqref{aequatio}. The resulting vacuum energy, which is positive and was first computed in~\cite{so16xso16}, indicates that the system exerts a back-reaction on space time. Its net result at the torus level is the emergence of a string--scale space--time curvature ${\cal O}(1/\alpha^\prime)$: this invalidates the original description around a Minkowski background, and we have no better tool yet, in general, to investigate the fate of this class of vacua than the low--energy (super)gravity.

In order to complete our presentation of non--tachyonic models in ten dimensions, it is now necessary to digress briefly from our main theme, and to address briefly two tachyonic models of oriented closed strings~\cite{type0}. They are usually called $0A$ and $0B$ strings, and their partition functions read
\bea
{\cal T}_{0A} &=& \int_{\cal F} \ \frac{d^2 \tau}{(Im \tau)^2} \ \frac{|O_8|^2 + |V_8|^2 + S_8\,\bar{C}_8+ S_8\,\bar{C}_8}{(Im \tau)^4 \ \eta^8 \, \bar{\eta}^8}\ , \nonumber \\
{\cal T}_{0B} &=& \int_{\cal F} \ \frac{d^2 \tau}{(Im \tau)^2} \ \frac{|O_8|^2 + |V_8|^2+|S_8|^2 + |C_8|^2}{(Im \tau)^4 \ \eta^8 \, \bar{\eta}^8} \ , \label{0A_0B}
\eea
so that their spectra are purely bosonic. Their low--lying modes comprise a tachyon $T$ and the other massless modes below:
\beq { 0A: \ ( e_\mu^a,B_{\mu\nu}, \phi, A_\mu^{1,2}, C_{\mu\nu\rho}^{1,2})}\ , \qquad {0B: \ ( e_\mu^a,B_{\mu\nu}^{1,2,3},\phi^{1,2,3},D_{\mu\nu\rho\sigma})}\ . \label{0AB_spectra}
\eeq

The presence of the tachyon is signalled by the first contribution to either of these partition functions, which involves the $O_8$ character in isolation. One of their open descendants, however, is the tachyon--free $U(32)$ model of~\cite{u32}, which we shall illustrate shortly. Before moving to orientifolds, however, let us describe a link between type $IIA$ and type $IIB$ that is purely ten--dimensional, in contrast with the $T$--duality upon circle compactification that is usually stressed. An orbifold construction which turns one of these models into the other can be induced if one attempts to project the $IIB$ spectrum by the operation $(-1)^{F_R}$, where $F_R$ is the fermion number of right--moving excitations. This projection amounts to replacing the original $IIB$ partition function in eq.~\eqref{IIA_IIB} with
\beq
 {{\cal T}_{IIB} \ \rightarrow \ \frac{1}{2}\ {\cal T}_{IIB}\ +\ \frac{1}{2} \ \int_{\cal F} \ \frac{d^2 \tau}{(Im \tau)^2} \ \frac{\left(V_8 \,-\,S_8\right)\left({\overline V}_8 \,+\,{\overline S}_8 \right)}{(Im \tau)^4 \ \eta^8 \, \bar{\eta}^8} } \ , \label{T_Untw}
 \eeq
but now the second term is not invariant under the modular transformation corresponding to the matrix $S$ in eq.~\eqref{modular}. The remedy is to add to eq.~\eqref{T_Untw} the term thus generated,
\beq
 {\frac{1}{2} \ \int_{\cal F} \ \frac{d^2 \tau}{(Im \tau)^2} \ \frac{\left(V_8 \,-\,S_8\right)\left({\overline O}_8 \,-\,{\overline C}_8 \right)}{(Im \tau)^4 \ \eta^8 \, \bar{\eta}^8} } \ , \label{T_tw1}
 \eeq
together with another obtained from it by a $T$ transformation,
\beq
{\frac{1}{2} \ \int_{\cal F} \ \frac{d^2 \tau}{(Im \tau)^2} \ \frac{\left(V_8 \,-\,S_8\right)\left(\,-\,{\overline O}_8 \,-\,{\overline C}_8 \right)}{(Im \tau)^4 \ \eta^8 \, \bar{\eta}^8} } \ . \label{T_tw2}
\eeq
Putting together the terms in eqs.~\eqref{T_Untw}, \eqref{T_tw1} and \eqref{T_tw2} does build a modular invariant result, but this is precisely the partition function of type $IIA$ in eq.~\eqref{IIA_IIB}. In other words, as we had anticipated, $IIA$ is an orbifold of $IIB$ (and vice versa).

Let us now illustrate how open--string models emerge as descendants (or orientifolds) of the left--right symmetric closed models that we have described, following~\cite{orientifolds}. The procedure is reminiscent of the orbifold that we have described, but it is more complicated since it also affects the two--dimensional surfaces, and thus the very nature of strings. Let us begin by illustrating it for the type--$I$ superstring. The starting point is in this case the type--$IIB$ model, which is symmetric under the interchange of left and right modes, as we have remarked. The operation begins with the addition of a Klein--bottle term to the (halved) torus partition function, a step which is reminiscent of the projection in eq.~\eqref{T_Untw}, so that
\beq
{\cal T}_{IIB} \ \to \ \frac{1}{2} \  {\cal T}_{IIB} \ + \ \frac{1}{2} \ \int_0^\infty \ \frac{d\tau_2}{(\tau_2)^2} \ \frac{V_8 \,-\,S_8}{(\tau_2)^4 \ \eta^8} \ [2 i \tau_2] \ .
\eeq
The last term is the ``direct--channel'' Klein--bottle amplitude, and the argument of the functions involved is indicated within square brackets. It completes the (anti)symmetrization, so that, say, starting from the $8 \times 8$ states present in the massless $NS-NS$ or $R-R$ sectors of the original oriented closed strings, one is left with 36 states in the former and 28 states in the latter, combinations that are respectively symmetric and antisymmetric under the interchange of left and right oscillators, as determined by the signs of the corresponding contributions to ${\cal K}$. In this fashion, the massless $NS-NS$ sector looses the two--form, while the massless $R-R$ sector looses the scalar and the self--dual four--form. The fermionic terms are simply halved, so that only one combination of the two gravitini and only one combination of the two type--$IIB$ spinors remains.

Notice that the Klein--bottle amplitude is not invariant under modular transformations. It exhibits nonetheless an interesting behavior if $\tau_2$ is halved, which amounts to referring the measure to its doubly--covering torus, and then an $S$ transformation is performed. This turns the second contribution into a \emph{tree--level} exchange diagram for the closed string spectrum between a pair of crosscaps (real projective planes, or if you will spheres with opposite points identified). The end result reads
\beq
{\widetilde {\cal K}} \ = \ \frac{2^5}{2} \ \int_0^\infty \ {d\ell} \ \frac{V_8 \,-\,S_8}{\eta^8} \ [i \ell] \ , \label{Ktilde}
\eeq
where the argument of the functions involved is again within square brackets.

Notice that all powers of $\ell$ have disappeared, as pertains to such a vacuum exchange, since they would signal momentum flow. Now this ${\widetilde {\cal K}}$ amplitude is akin, in many respects, to the other two possible types of \emph{tree--level} exchange diagrams, those between a pair of boundaries and between a boundary and a crosscap, ${\widetilde{\cal A}}$ and ${\widetilde{\cal M}}$,
\beq
{\widetilde {\cal A}} \,=\, \frac{2^{-5}}{2} \ {\cal N}^2 \ \int_0^\infty \ {d\ell} \ \frac{V_8 \,-\,S_8}{\eta^8} \ [i \ell] \ , \quad
{\widetilde {\cal M}} \,=\, - \ 2\ \frac{1}{2} \ {\cal N} \ \int_0^\infty \ {d\ell} \ \frac{V_8 \,-\,S_8}{\eta^8} \ [i \ell+ 1/2] \ . \label{AMtilde}
\eeq
Both expressions involve the same closed spectrum, but the reader should appreciate a few facts. The first is the presence of a Chan--Paton multiplicity~\cite{chanpaton} ${\cal N}$ associated to each boundary, which thus enters quadratically the first amplitude and linearly the second. The second fact is the presence of a shifted argument in the second contribution ${\widetilde{\cal M}}$, consistently with its skew doubly covering torus. The other relevant ingredient is the combinatoric factor of two present in the second expression, while its overall ``minus'' sign guarantees that the overall contribution to the $S_8$ sector, proportional to
\beq
\frac{2^5}{2} \left( 1 + {2^{-5}\, \cal N}^2 \ - \ 2\, \times \, 2^{-5}\, \cal N\right) \ , \label{tadpole}
\eeq
add up to zero for ${\cal N}=32$. The vanishing of the massless ``tadpole'' term associated to $S_8$ lies at the heart of the Green--Schwarz anomaly cancellation~\cite{greenschwarz}, as a crucial prerequisite, since it signals the elimination of the irreducible sixth--order contribution to the twelve--dimensional anomaly polynomials. Notice, however, that supersymmetry implies that the corresponding $V_8$ term also cancels: we shall soon explain what happens when this is not the case.

One can now perform two distinct inverse operations that are the counterparts of the one that led to eq.~\eqref{Ktilde}. The first involves the matrix $S$ and then a rescaling $\tau_2 \to \tau_2/2$, which turns the transverse annulus amplitude ${\widetilde {\cal A}}$ into its direct--channel form ${\cal A}$. The second involves the action on ${\widetilde {\cal M}}$ with the combination
\beq
P=T^\frac{1}{2}\,S\,T^2\,S\,T^\frac{1}{2} \label{Pmat}
\eeq
after rescaling $\ell \to \ell/2$, which turn the transverse M\"obius amplitude ${\widetilde {\cal M}}$ into its direct--channel form ${\cal M}$. The fractional powers of $T$ in eq.~\eqref{Pmat} stem from a convenient redefinition of the functions involved in the M\"obius contribution, which keeps the (rescaled) ``hatted'' contributions real despite the presence of a real part of $\tau$ in their arguments. The end result is the open spectrum with the proper Regge slope,
\bea
{{\cal A}} &=& \frac{1}{2} \ {\cal N}^2 \ \int_0^\infty \ \frac{d\tau_2}{(\tau_2)^2} \ \frac{V_8 \,-\,S_8}{(\tau_2)^4 \ \eta^8} \ [ i \tau_2/2] \ , \nonumber \\
{{\cal M}} &=& - \ \frac{1}{2} \ {\cal N} \ \int_0^\infty \ \frac{d\tau_2}{(\tau_2)^2} \ \frac{{\hat V}_8 \,-\,{\hat S}_8}{(\tau_2)^4 \ {{\hat \eta}^8}} \ [i \tau_2/2 + 1/2] \ , \label{AM}
\eea
as manifested by the arguments of these expressions, which are indicated within square brackets. There are $\frac{{\cal N}({\cal N} - 1)}{2}$ massless gauge bosons and gaugini, and thus of a ten--dimensional supersymmetric Yang--Mills theory with an $SO(32)$ group, in addition to a closed sector corresponding to the type--$I$ (or $(1,0)$) supergravity, as in the HE and HO cases, but for an important difference: here the two--form comes from the Ramond--Ramond sector. In Polchinski's space--time picture~\cite{polchinski}, boundaries and crosscaps translate into $D$-branes and orientifold planes in spacetime, so that the steps that we have described translate into the introduction of a ten--dimensional $O_-$ orientifold~\footnote{Note that this convention is opposite to the one of~\cite{wittenOplus} and of the first review in~\cite{orientifolds}, where $O_\mp$ would be called $O_\pm$.}, with negative tension and charge, and of a number of $D$--branes. Given that $D$--branes have by definition a positive tension and charge, the corresponding signs for the orientifold can be read from the coefficients of $V_8$ and $-S_8$ in the vacuum--channel M\"obius contribution.

The three additional contributions appear to suffer individually from an ultraviolet problem, but in fact the $S$ and $P$ transformations turn them into \emph{tree--level} vacuum exchange amplitudes, as we have stressed, converting at the same time the ultraviolet region into an infrared one. Hence, the ultraviolet problem is again absent, while the infrared behavior is again potentially subtle and rich.

The Sugimoto model~\cite{sugimoto} is apparently a minor variant of this construction, so much so that it was long overlooked. Our own involvement with the physical phenomenon it entails, brane supersymmetry breaking, started actually from some more complicated manifestations of it that showed up early, in attempts made by one of us with Massimo Bianchi and Gianfranco Pradisi, which are summarized in~\cite{bsb_old}. Surprisingly some tadpole conditions, more complicated counterparts of eq.~\eqref{tadpole}, seemed to yield inconsistent solutions. This result was also in conflict with the apparent supersymmetry of the models, and so the whole story remained a puzzle for long, until its origin clarified in~\cite{bsb}, for the whole perturbative spectrum of six--dimensional models involving different types of branes. The clear identification, in~\cite{wittenOplus}, of the $O_\pm$ orientifold planes, which had entered the rank--reducing toroidal compactifications in~\cite{orientifolds}, was also an important ingredient, but Sugimoto's construction remains the simplest example of this type, since it involves a single type of brane and orientifold. The only change, in it, with respect to the supersymmetric $SO(32)$, is the sign of the $V_8$ contribution to ${\cal M}$, which now reads
\beq
{{\cal M}} \ = \  - \ \frac{1}{2} \ {\cal N} \ \int_0^\infty \ \frac{d\tau_2}{(\tau_2)^2} \ \frac{ -\, V_8 \,-\,S_8}{(\tau_2)^4 \ \eta^8} \ [i \tau_2/2 + 1/2] \ .
\eeq
The effect of this simple operation, however, is dramatic: there are now $\frac{{\cal N}({\cal N} + 1)}{2}$
gauge bosons, while the anomaly cancellation, signalled by the Ramond--Ramond contribution $-S_8$  to the vacuum channel, continues to require the presence of $\frac{{\cal N}({\cal N} - 1)}{2}$ Fermi fields with ${\cal N}=32$. All in all, one gets a $USp(32)$ gauge group, but the fermions remain in the antisymmetric representation, which is reducible in $USp(32)$ and contains a singlet. The singlet is most important in the overall picture, since it is the goldstino. \emph{Supersymmetry is broken, and is non--linearly realized here, without an order parameter capable of recovering it}, which is a startling phenomenon to occur in String Theory. In the space--time picture, this novelty is induced since the $O_-$ orientifold is replaced by an $O_+$, with positive tension and charge. The positive charge requires anti $D$--branes for its cancelation, and a net tension remains in the vacuum. As a result, here we face a more intricate example of the ultraviolet--infrared correspondence, and matters are to be dealt with carefully. The ${\cal M}$ amplitude appears ultraviolet divergent, but in ${\widetilde{\cal M}}$ the effect reveals its infrared origin from a massless exchange at zero momentum, and therefore from the presence of a (projective)disk $NS-NS$ tadpole to which a (singular) massless propagator of zero momentum can couple. The way out is to shift to another vacuum which solves the equations of motion~\cite{vacuum_redefinitions}, but as of today this can be done efficiently only at the level of the low--energy effective field theory. While the key question concerns vacuum configurations, one should also address a simpler problem, which is to characterize how the (charged or uncharged~\cite{sen}) $D$-branes available in these systems~\cite{dms} adjust themselves to the deformed backgrounds.
\begin{figure}[ht]
\centering
\includegraphics[width=80mm]{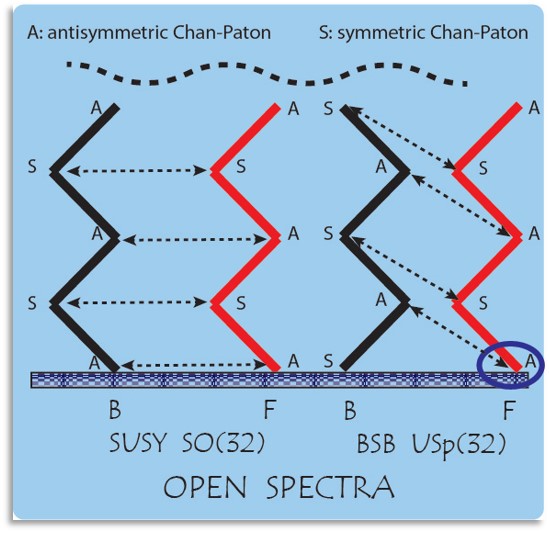}
\caption{\small ``Brane supersymmetry breaking'' induces a shift by one level of Bose excitations with respect to the supersymmetric type--$I$ open spectrum, with a minimal Bose--Fermi unpairing.}
\label{fig:bsb_spectrum}
\end{figure}

Supersymmetry is broken in this $USp(32)$ model due to the simultaneous presence, in the vacuum, of objects that are individually BPS but not mutually so. The scale of the breaking can be estimated from the cartoon in fig.~\ref{fig:bsb_spectrum} as the string scale, which makes one wonder again how much the effective low--energy supergravity, which is the only tool at our disposal, can possibly deal with it. For one matter, once clearly misses the relatively ``soft'' nature of the phenomenon, which becomes apparent if one considers the whole spectrum as represented in fig.~\ref{fig:bsb_spectrum}. Notice that, in comparing with the $SO(32)$ superstring, the whole ``belt'' of bosonic states has merely moved around by one step.  The contribution to the vacuum energy can be manifested directly making use of Jacobi's \emph{aequatio} of eq.~\eqref{aequatio}, but it is alas infinite! The infinity, however, can be ascribed to an \emph{infrared} effect, the emergence of \emph{a finite new interaction that couples to a massless propagator of zero momentum}. In some respect, the phenomenon is thus simpler than the quantum fluctuations that were present in the heterotic $SO(16)\times SO(16)$, which makes coming to terms with it ever more urgent. The new contribution to the $USp(32)$ model is a runaway dilaton potential, which we shall often refer to as a ``dilaton tadpole''. The resulting ``string frame'' effective action reads
\bea
{\cal S} &=& \frac{1}{2\,k_{10}^2}\ \int d^{10}x\,\sqrt{-\,G}\bigg\{ e^{-2\phi}\left[\, - \,R\, + \,4(\partial\phi)^2 \right]\nonumber \\ &-& \frac{1}{2(p+2)!}\ e^{-2\,\beta_S \,\phi}\ {\cal H}_{p+2}^2  \ -\ T \ e^{\,\gamma_S\,\phi} \, \bigg\} \ , \label{string_frame_bsb}
\eea
with $p=1$, $\beta_S=0$ and $\gamma_S=-1$. A term similar to the last one, also with a positive overall coefficient, would be induced in the heterotic $SO(16) \times SO(16)$ model at the torus level, and therefore with $\gamma_S=0$, while the $NS$ nature of its two--form field ($p=1$, again), would imply that $\beta_S=1$. The Weyl rescaling $G \to g \,e^{\phi/2}$ turns the action \eqref{string_frame_bsb} into its Einstein--frame form
\beq
{\cal S} \,=\, \frac{1}{2\,k_{10}^2}\, \int d^{10}x\sqrt{-g}\bigl\{-\ R\ - \frac{1}{2}\ (\partial\phi)^2 \,-\,  \frac{1}{2\,(p+2)!}\ e^{-2\,\beta_E^{(p)}\,\phi}\, {\cal H}_{p+2}^2 \ - \ T \, e^{\,\gamma_E\,\phi} \label{einstein_frame_bsb}
\bigr\} \ ,
\eeq
where now $\gamma_E=\frac{3}{2}$ for the orientifold model and $\gamma_E=\frac{5}{2}$ for the heterotic $SO(16) \times SO(16)$ model, while $\beta_E=-\frac{1}{2}$ for the orientifold model and $\beta_E=\frac{1}{2}$ for the heterotic $SO(16) \times SO(16)$ model.

There is actually another tachyon--free model in ten dimensions~\cite{u32}, whose applications have been pursued in~\cite{armoni}. It is a peculiar orientifold of the tachyonic $0B$ model, obtained starting from the Klein--bottle amplitude
\beq
{\cal K}_{0'B} \ = \ \frac{1}{2} \ \int_0^\infty \ \frac{d\tau_2}{(\tau_2)^2} \ \frac{\,-\,O_8\,+\,V_8 \,+\,S_8\,-\,C_8}{(\tau_2)^4 \ \eta^8} \ [2 i \tau_2] \ , \label{K0bprime}
\eeq
which is indeed designed to project out the closed--string tachyon. The choice of signs determines a pattern of (anti)symmetrizations that is consistent with the interactions of the various string sectors, or if you will with the ``fusion rules'' of the Conformal Field Theory. There are other tachyonic descendants of
the 0A and 0B models, all with fermions in the open sector, which were introduced in the 1990 PLB of~\cite{orientifolds}, and another variant was also introduced in~\cite{u32}.

Even at the cost of indulging on old activity, we cannot resist mentioning here that this construction was an outgrowth of work done by one of us with a dear friend and collaborator, Yassen Stanev, who is deeply missed (and with Pradisi)~\cite{pss}, where the two--dimensional consistency condition for crosscaps of~\cite{crosscap_constraint}, obtained extending the results in~\cite{lewellen} to non--orientable surfaces, was generalized (these results were further extended in~\cite{schellekens}). A direct investigation of the open descendants of WZW~\cite{WZW} models showed in fact that their different sectors allowed multiple choices for ${\cal K}$ that determined, in their turn, corresponding choices for ${\cal A}$ and ${\cal M}$. This early counterpart of the $O_\pm$ option underlies indeed the peculiar Klein bottle amplitude of eq.~\eqref{K0bprime}. In the ten--dimensional models, the Minkowski fusion rules imply, for example, that $O_8 \times O_8$ gives $V_8$, or if you will that the mutual interactions of two states in the $O_8$ sector give rise to state in the $V_8$ sector, so that one can allow for antisymmetrized states in the $O_8$ sector compatibly with the inevitable symmetrization of the $V_8$ sector, which yields graviton and dilaton modes, and so on. These properties are directly implied by the Verlinde formula~\cite{verlinde} for the $(O_8,V_8,-S_8,-C_8)$ Minkowski sectors. The open spectrum then follows letting in principle the allowed sectors, which here are all the four available ones, flow in the vacuum exchanges. There is a novel subtlety, however, since this model involves ``complex'' charges, and thus a unitary $U(32)$ gauge group. Its annulus and M\"obius amplitudes read
\bea
{\cal A}_{0'B} &=&  \int_0^\infty \ \frac{d\tau_2}{(\tau_2)^2} \ \frac{{ {\cal N}} \, { {\cal {\overline N}}} \ V_8 \,-\, \frac{1}{2} \ ( {\cal N}^2 \,+\,{\cal {\overline N}}^2 ) \ C_8}{(\tau_2)^4 \ \eta^8} \ [i \tau_2/2] \ , \nonumber \\
{\cal M}_{0'B} &=& - \ \frac{{ {\cal N}\,+\, {\cal {\overline N}}}}{2}\ \int_0^\infty \ \frac{d\tau_2}{(\tau_2)^2} \ \frac{\hat{C}_8}{(\tau_2)^4 \ \hat{\eta}^8} \ [i \tau_2/2+1/2] \ ,
\eea
while the corresponding exchange amplitudes, which read
\bea
{\widetilde{\cal K}}_{0'B} &=&  - \ \frac{2^6}{2} \ \int_0^\infty \ {d\ell} \ \frac{C_8}{ \eta^8} \ [i \ell] \ , \nonumber \\
{\widetilde{\cal A}}_{0'B} &=&  \frac{2^{-6}}{2} \ \int_0^\infty \ {d\ell} \ \frac{({\cal N}+{\overline{\cal N}})^2 \,(V_8 - C_8) \ - \ ({\cal N}-{\overline{\cal N}})^2 \,(O_8 - S_8)}{ \eta^8} \ [i \ell] \ , \nonumber \\
{\widetilde{\cal M}}_{0'B} &=& - \ 2\ \frac{{ {\cal N}\,+\, {\cal {\overline N}}}}{2}\ \int_0^\infty \ {d\ell} \  \frac{\hat{C}_8}{ \hat{\eta}^8} \ [i \ell+1/2]
\eea
are consistent thanks to the constraint ${\cal N}= {\overline{\cal N}}$, which reflects the numerical coincidence of the dimensions of the fundamental $U(32)$ representation and its conjugate and eliminates the $S_8$ contribution. As in preceding examples with broken supersymmetry or without supersymmetry altogether, the system exerts a back--reaction on spacetime, in this case both via the torus amplitude and again via the mere insertion in the vacuum of branes and orientifolds, which here are not BPS to begin with.

There is an important difference between the effects of supersymmetry breaking in the torus amplitude and in the open and unoriented sectors, which is worth stressing. The former is a genuine quantum effect, and in principle one could also look for compensation mechanisms among different mass levels that make the resulting effects milder~\cite{vanishing_nonsusy}, while the latter reflects the mutual attraction of extended objects that populate the vacuum, and in this respect it is a ``harder'' effect. Still, branes and antibranes can move, and their mutual attraction can lead to tachyonic instabilities. In general, it is very difficult to concoct equilibrium configurations, even in the presence of special symmetries~\cite{kita_17}. On the other hand, orientifolds are not dynamical, and the resulting beauty of brane supersymmetry breaking is precisely that no tachyons are introduced. Nonetheless, the mutual attraction between branes and orientifolds makes one shiver a bit, since the resulting spacetimes would seem prone to collapsing.
As we shall see shortly,  these systems seem to perform well nonetheless at least in the evolving backgrounds of Cosmology, where they can also convey some interesting and surprising lessons.

\section{Classical solutions in the presence of a dilaton tadpole}\label{sec:classical_sols}

In the preceding section we have highlighted the key difficulty introduced by brane supersymmetry breaking, the residual attractive force between the extended objects that populate the vacuum. In this section we shall see that, despite all this, the system affords an interesting cosmological behavior, with potential lessons for the onset of inflation. Static backgrounds, however, do present the expected difficulties and, as of today, they are still awaiting a convincing treatment. We shall actually consider, at the same time, the $USp(32)$ model with brane supersymmetry breaking and the $U(32)$ model, whose low--energy Lagrangians, when restricted to the sector relevant to the vacuum configurations at stake, coincide, and the $SO(16) \times SO(16)$ model, where the dilaton potential induced by the torus correction is slightly different but has similar effects. The fact that two very different models like the $USp(32)$ and $U(32)$ strings look the same from this vantage point is a reminder, if need there be, of the limitations inherent in the low--energy approach.

With this proviso, we shall take as our starting point the generic string--frame action of eq.~\eqref{string_frame_bsb}, or its Einstein--frame counterpart of eq.~\eqref{einstein_frame_bsb}. This is all one can do in a handy fashion, but it would be clearly interesting to account in a more satisfactory way for the pattern of massive excitations, where the Fermi--Bose pairing is essentially recovered, up ${\cal O}(\alpha^\prime)$ differences, in the $USp(32)$ model of brane supersymmetry breaking.

The first classical solution of these models was presented in~\cite{dm_9Dsolution} by Emilian Dudas and one us. It describes a compactification to nine dimensions on an interval, which is of finite length in the resulting metric. In the Einstein frame, the solution for the $USp(32)$ and $U(32)$ orientifold models reads
\begin{eqnarray}
&& ds^2 \ = \ \left|\alpha\,y^2\right|^\frac{2}{9}\,e^{\frac{1}{2}\,\phi_0} \, e^{\,\frac{1}{4}\,\alpha\,y^2}  \, \eta_{\mu\nu} \, dx^\mu\,dx^\nu \ + \ \left|\alpha\,y^2\right|^{\,-\,\frac{1}{3}}\,e^{\,-\,\phi_0} \, e^{\,-\,\frac{3}{4}\,\alpha\,y^2} \, dy^2 \ , \nonumber \\
&& e^\phi \ = \ e^{\phi_0}  \left| \alpha\,y^2\right|^\frac{1}{3} \ e^{\,\frac{3}{4}\,\alpha\,y^2}
\ . \label{9D_solution}
\end{eqnarray}
This result has the interesting property of leading to a flat nine--dimensional space--time with finite values of nine--dimensional Newton constant and gauge coupling, but presents two annoying features. To begin with, the reader will notice that the string coupling in the second equation grows indefinitely for $y \to + \infty$. Moreover, the internal curvature scale grows indefinitely at the opposite end, so that the result is expected to receive large $\alpha^\prime$ and string--loop corrections. A similar behavior obtains in a corresponding solution for the $SO(16) \times SO(16)$ model.

The situation improves in the corresponding cosmological solution,
\begin{eqnarray}
&& ds^2 \ = \ \left|\alpha\,t^2\right|^\frac{2}{9}\,e^{\frac{1}{2}\,\phi_0} \, e^{\,-\,\frac{1}{4}\,\alpha\,t^2}  \, \delta_{ij} \, dx^i\,dx^j \ - \ \left|\alpha\,t^2\right|^{\,-\,\frac{1}{3}}\,e^{\,-\,\phi_0} \, e^{\,\frac{3}{4}\,\alpha\,t^2} \, dt^2 \nonumber \\
&& e^\phi \ = \ e^{\phi_0}  \left| \alpha\,t^2\right|^\frac{1}{3} \ e^{\,-\,\frac{3}{4}\,\alpha\,t^2}
\ .
\end{eqnarray}
which was also first obtained in~\cite{dm_9Dsolution}.
The key difference has to do with the gaussian cut in the behavior of the string coupling for large times, which is induced by the analytic continuation $y \to it$: there is now an upper bound on $\phi$, and thus on the string coupling! Notice however that the two ends of the evolution, $t=0$ and $t=\infty$, still host curvature singularities that signal the need for large $\alpha'$ string corrections. While one may not worry about the latter, since this is at most a model for Early Cosmology, the ensuing discussion remains qualitative, to some extent, since the former singularity is indeed present.

The study of exponential potentials in inflationary Cosmology has a long history~\cite{cosmo_power}, while the result in~\cite{dm_9Dsolution} was extended in~\cite{russo} to general exponential potentials of the type
\beq
V \ = \ T \, e^{\gamma\,\phi} \ . \label{climbing_pot}
\eeq
Interestingly, the case corresponding to brane supersymmetry breaking in ten dimensions, $\gamma=\frac{3}{2}$, and up to an assumption that was spelled out in~\cite{integrable}, its lower--dimensional counterparts, all correspond to a ``critical behavior'' separating two different dynamical regimes. Another scalar mode enters indeed compactifications to lower dimensions, the one parametrizing the internal volume, and mixes with the dilaton. The end result in quite interesting, since one combination of the two retains a critical potential for all lower dimensions $D$. Hence, up to the stabilization of the second scalar, which could be induced by some other mechanism, the ``critical'' behavior is universally present.
The resulting physical picture is enticing, and was clarified in~\cite{climbing}: for low enough values of $\gamma$, which one can assume to be positive up to a field redefinition, the scalar can ``emerge'' from the initial singularity from the right, descending the potential, or from the left, climbing it~\footnote{Climbing also occurs in the $SO(16) \times SO(16)$ model, whose ``hypercritical'' potential arises from the torus amplitude.}. As $\gamma$ increases, however, one soon encounters a critical value $\gamma_{crit}$, which is precisely $\frac{3}{2}$ in ten dimensions and would be $\sqrt{6}$ in four dimensions, where the behavior changes drastically. The descending solution disappears for $\gamma \geq \gamma_{crit}$ and the only option left for $\phi$ (or for the proper combination) is to climb the potential up to a certain height, to then revert its motion and start a descent. In String Theory $e^\phi$ is the coupling that sizes the loop expansion, so that the climbing behavior is potentially under control in perturbation theory. This is not the whole story, of course, since close to the initial singularity curvature corrections become very large, but one is somehow half of the way in the right direction. Other interesting solutions for the potential of eq.~\eqref{climbing_pot} were discussed in~\cite{cd}, where some puzzles related to this dynamics are also addressed.

The ``climbing'' picture~\cite{climbing} has counterparts in a wide class of integrable cosmologies~\cite{integrable}, and suggests very naturally a mechanism to start inflation~\cite{inflation}, provided one identifies a proper combination of dilaton and internal volume mode with the inflaton: a fast inflaton compelled to climb would reach a turning point, stopping there momentarily before descending and attaining slow--roll as it releases its original energy. String Theory gives at present no indications on the emergence of a properly flat portion of the potential, but if this were the case inflaton deceleration could leave, in principle, some detectable signatures. The target of these considerations would be the observed low--$k$ depression of the CMB angular power spectrum, which seems to deviate from the $\Lambda$CDM concordance model~\cite{WMAP,PLANCK}. If this effect were not a mere fluctuation, a decelerating inflaton could account for a depression of primordial power spectra, which a short--enough inflation could have imprinted, almost verbatim, on low--$\ell$ CMB multipoles~\cite{depression}. To reiterate, brane supersymmetry breaking leads naturally to envisage scenarios of inflaton deceleration, and \emph{deceleration provides a natural scenario to induce a low--$\ell$ lack of power in the CMB angular power spectrum}.

One can play with potentials including a ``critical'' term and try to optimize $\chi^2$ fits of the low--$\ell$ CMB multipoles ascertained by WMAP~\cite{WMAP} and {\sc Planck}~\cite{PLANCK} for $\ell \leq 30$, as in~\cite{low_l_fits}. This is admittedly a theoretical game, and the oscillations are likely to be mere fluctuations, but a few facts resist a more robust analysis that we are about to review:
\begin{itemize}
\begin{figure}[ht]
\centering
\includegraphics[width=65mm]{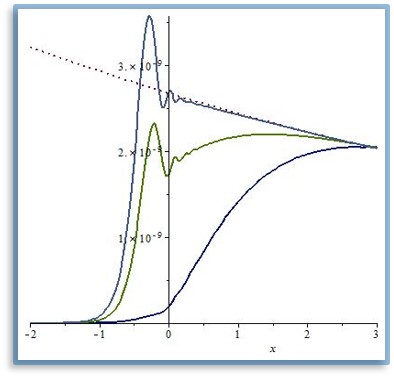}
\caption{\small Some primordial power spectra obtained from a double exponential potential of the type $V=\exp{(\sqrt{6}\phi)}+\exp{(\sqrt{6}\,\gamma\,\phi)}$, with $\gamma \simeq 1/12$: the three curves, from top to bottom, correspond to progressively harder impacts of $\phi$ on the exponential barrier described by the first term. }
\label{fig:2exp_spectra}
\end{figure}
\item at low Galactic latitudes the type of behavior associated to a climbing scalar, with a peak in the primordial power spectrum that precedes a Chibisov--Mukhanov~\cite{CM} plateau, as in the intermediate plot of fig.~\ref{fig:2exp_spectra}, appears slightly favored. It is generally present in potentials that combine a critical exponential with milder terms allowing for an inflationary epoch, and the addition of a second, milder exponential, can model a plateau and suffices to exhibit it. Of course, these are simple toy models and lead, as is well known, to a tensor--to--scalar ratio that is way too large;
\begin{figure}[ht]
\centering
\includegraphics[width=65mm]{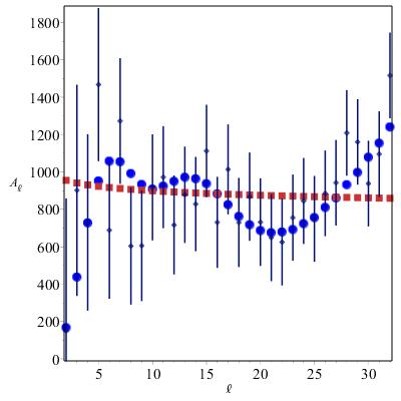}
\caption{\small An optimal fit obtained in~\cite{low_l_fits} of low--$\ell$ CMB multipoles with minimal masking, which leaves about 93\% of the CMB sky, obtained adding to a double exponential potential $V=\exp{(\sqrt{6}\phi)}+\exp{(\sqrt{6}\,\gamma\,\phi)}$ a gaussian defect in the transition region between the two exponentials.}
\label{fig:narrowmask}
\end{figure}
\item one can also arrive at an amusingly close correspondence with the pattern of oscillations present for $\ell <30$, combining the preceding potential with a small gaussian defect around the ``critical'' exponential wall~\cite{low_l_fits} that slows down the reversal of the scalar motion (see fig.~\ref{fig:narrowmask});
\begin{figure}[ht]
\centering
\includegraphics[width=65mm]{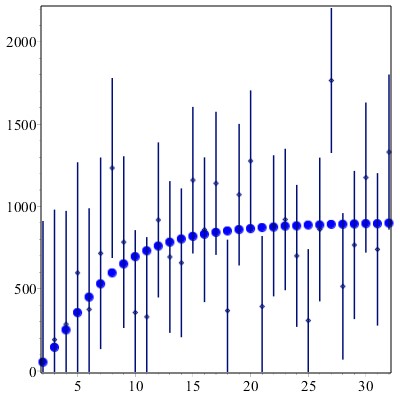}
\caption{\small An optimal fit of low--$\ell$ CMB multipoles~\cite{erice15} to the universal cut of~\cite{depression}, with a wider Galactic mask that leaves about 39\% of the CMB sky, where oscillations fade out to a large extent.}
\label{fig:widemask}
\end{figure}
\item the oscillations in the first CMB multipoles fade out to a large extent at higher Galactic latitudes (see fig.~\ref{fig:widemask}), where~\cite{erice15} the overall pattern is well captured by the low--$k$ modification of the primordial power spectrum described in~\cite{depression}:
    \beq
    P_S \ \simeq \  A \ k^{n_s -1} \quad \longrightarrow \quad P_S \ \simeq \ A \ \frac{k^3}{\left[k^2 \ + \ \Delta^2\right]^{2-\frac{n_s}{2}}} \ . \label{power_cut}
    \eeq
\end{itemize}

The simple expression in eq.~\eqref{power_cut} captures the universal low--frequency cut present in all models of inflaton deceleration. It obtains subjecting the standard inflationary Mukhanov--Sasaki inflationary potential $W \sim 1/\eta^2$, where $\eta$ denotes conformal time, to a vertical shift that brings its past end below the real axis, so that the negative portion of $W$ models the result of an early fast--roll. The region hosting the cut terminates around the scale $\Delta$, where the primordial power spectrum begins to converge toward the standard Chibisov--Mukhanov tilt~\cite{CM} also displayed in eq.~\eqref{power_cut}. As is well known, the spectral index in this key expression was ascertained to high precision by the {\sc Planck} collaboration~\cite{PLANCK}.
\begin{figure}[ht]
\centering
\includegraphics[width=65mm]{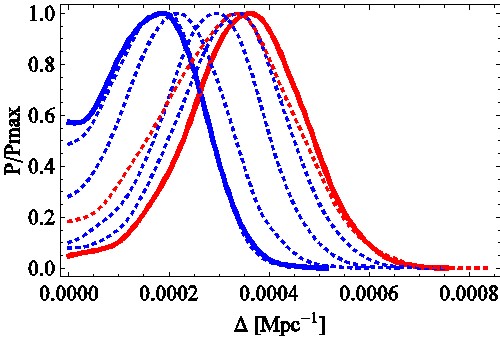}
\caption{\small Posteriors for the parameter $\Delta$ in eq.~\eqref{power_cut} with different maskings around the Galactic plane. The color coding is as follows:  solid blue for the 94\% mask, thick red for a $+30^\circ$ extension, dotted blue for the intermediate masks with $+6^\circ$, $+12^\circ$, $+18^\circ$ and $+24^\circ$, dotted red for $36^\circ$.}
\label{fig:delta}
\end{figure}

One can actually play a more sophisticated game and extend the standard concordance model $\Lambda$CDM into what might be called $\Lambda$CDM$\Delta$, while attempting a determination of the scale $\Delta$ in eq.~\eqref{power_cut}. This was done in~\cite{delta_cut}, and the simple analysis in~\cite{erice15} resonates with the end result of the detailed investigation: higher Galactic latitudes favor a determination of $\Delta$, whose detection level rises to about 3 $\sigma$ in a mask obtained by a $30^\circ$ blind extension of the minimal {\sc Planck} mask (see fig.~\ref{fig:delta}). Notice that one is thus left with about 39\% of the CMB sky, to be compared with the 94\% of it that would be allowed by the minimal {\sc Planck} mask. The CMB might be cleaner well away from the Galactic plane, so that there might be something to this improved determination. At any rate the end result,
\beq
\Delta \ = \ \left(0.35 \ \pm \ 0.11\right) \times 10^{-3}\ {\rm Mpc}^{-1} \ ,
\eeq
where the error indicated corresponds to 68$\%$ C.L., translates into a distance scale that, as expected, is of the order of the Cosmic Horizon. In contrast, in the minimal {\sc Planck} mask one attains a lower detection level~\cite{delta_cut}, of order $88.5\%$, with $\Delta \ = \ \left(0.17 \ \pm \ 0.09\right) \times 10^{-3}\ {\rm Mpc}^{-1}$.

If the effect is really there and is not merely the result of a fluctuation, these findings could have interesting implications for Cosmology. To begin with, the new scale could signal the onset of inflation (or perhaps, more prosaically, a local feature of the inflationary potential), but it would then also impinge on structure formation, which ought to decrease at the very large scales affected by $\Delta$. This analysis was extended and improved more recently in~\cite{gklmns}: the upshot is that even multipoles behave very regularly and lead to a determination of $\Delta$ that is largely latitude independent, as would pertain to a cosmological parameter, while odd multipoles display a less regular behavior. It remains to be seen whether all this is merely another hint that the scale $\Delta$ is best detected in a subset of the present {\sc Planck} data, or, on the contrary, it reflects a fundamental property of the CMB pattern, whose origin remains to be ascertained. Following~\cite{depression}, in~\cite{gklmns} it is also stressed that, if the emergence of $\Delta$ is really connected to glimpses of a decelerating scalar in the early stages of inflation, the power cut in scalar and tensor temperature perturbations should result in an increase of the tensor--to--scalar ratio at low $\ell$ with a high tilt, albeit in a region where both signals would be reduced to some extent.

Summarizing, the simple critical potential of brane supersymmetry breaking has surely stimulated some constructive interactions with the Cosmology community, which were largely driven by some common interest in the apparent low--$\ell$ lack of power in the CMB. Time will tell where all this will lead. The better polarization measurements that the next generation of CMB experiments will allow can bring along far more precise determinations of $\Delta$ (or its disappearance altogether from the picture).

The situation becomes again intricate and less favourable if one tries a different, natural approach to static vacua, making use of (symmetric) internal fluxes. This type of approach has accompanied from the beginning, in different forms, the study of supersymmetric vacua in String Theory~\cite{CY}, and it was natural to explore it in this context as well~\cite{ms}. Once more, up to some minor differences one can do this, in an identical fashion, for the $USp(32)$ and $U(32)$ orientifold models, and also for the $SO(16) \times SO(16)$ heterotic model reviewed in Section~\ref{sec:10D}. Our starting point was provided by the class of metrics
\beq
ds^{\,2}\ = \ e^{2A(r)}\, g(\mathtt{L_k})\ + \ dr^2\ + \ e^{2C(r)}\, g(\mathtt{E_{k'}}) \ ,
\eeq
where $\mathtt{L_k}$ ($\mathtt{E_{k'}}$) is a maximally symmetric Lorentzian (Euclidean) spacetime with curvature $\mathtt{k} \, (\mathtt{k'})$, and by volume form fluxes of the type
\beq
{\cal H}_{p+2} \ = \ h\, e^{\,(p+1)\,A(r) \,+\, 2\, \beta_E^{(p)}\,\phi \, -\, (8-p) \,C} \ \epsilon(p+1) \, dr
\eeq
that invade the resulting spacetimes. The key requirement in~\cite{ms} was the need to avoid large string couplings, in contrast with the solution of eq.~\eqref{9D_solution}, which could be attained in principle requiring that the dilaton be stabilized. In this fashion we were led to vacua of the $AdS \times S$ type. Most interesting among these are vacua characterized by large sizes, which are expected to suppress string $\alpha'$ corrections, with also small values for the string coupling $g_s = \exp(\phi)$, which are expected to suppress string loop corrections. All in all, in a weakly--coupled corner of this type, the low--energy effective field theory would appear a reliable tool to gather some information on String Theory altogether. In~\cite{ms} we also allowed for an internal gauge field profile, but this was problematic~\cite{basile} and did not add to potentially interesting options, so that here we shall confine our attention to vacua where only a form profile is present.
Assuming also a constant dilaton profile, the vacuum equations reduce to
\begin{eqnarray}
T\, e^{\,\gamma_E\,\phi} &=& - \ \frac{\beta_E^{(p)}\,h^2}{\gamma_E} \ e^{\,-\,2\,(8\,-\,p)\,C\,+\,2\,\beta_E^{(p)}\,\phi} \ , \nonumber \\
16\,k'\, e^{\,-\,2\,C} &=& \frac{h^2 \left( p\ +\ 1 \, - \, \frac{2\,\beta_E^{(p)}}{\gamma_E} \right)e^{\,-\,2\,(8\,-\,p)\,C\,+\,2\,\beta_E^{(p)}\,\phi}}{\left(7\,-\,p\right)} \ , \nonumber \\
(A')^2 &=& k\,e^{\,-\,2\,A}
\ + \ \frac{h^2}{16(p\,+\,1)} \ \left( 7\,-\,p \,+ \, \frac{2\,\beta_E^{(p)}}{\gamma_E} \right)\ e^{\,-\,2\,(8\,-\,p)\,C\,+\,2\,\beta_E^{(p)}\,\phi}\ .
\end{eqnarray}
The first is the dilaton equation, which requires for consistency $\beta_E<0$, a condition that translates into the presence of a \emph{three--form} flux in the orientifold vacua and of a \emph{seven--form} flux in the heterotic $SO(16) \times SO(16)$ case. This fact determines the available options: $AdS_3 \times S^7$ in the orientifold case and $AdS_7 \times S^3$ in the heterotic case. On the other hand, the choice of one or another of the possible values of $k$, $0$ or $\pm 1$ selects different slicing of the same $AdS$ spacetime. The case $k=0$, in particular, captures $AdS$ in Poincar\'e -- like coordinates, and the detailed solutions read
\beq
g_s \ \equiv \ e^{\,\phi} \ = \ \frac{12}{\left(2\,h\,T^3\right)^\frac{1}{4}} \ , \qquad R^4  \, g_s^{3} \ = \ \frac{144}{T^2} \ , \qquad \left(A'\right)^2 \ = \ k \, e^{\,-\,2\,A} \ + \ \frac{{6}}{R^2} \label{ads3sol}
\eeq
for the orientifold case $(p=1)$ and
\beq
g_s \ \equiv \ e^{\,\phi} \ = \ \left(\frac{5}{h^2\,\Lambda^2}\right)^\frac{1}{4} \ , \qquad R^4  \, g_s^{5} \ = \ \frac{1}{\Lambda^2} \ , \qquad \left(A'\right)^2 \ = \ k \, e^{\,-\,2\,A} \ + \ \frac{{1}}{12\,R^2}
\eeq
for the heterotic $(p=5)$ case, where the one--loop contribution $\Lambda$, rather than $T$, accompanies the potential of eq.~\eqref{climbing_pot}.  Both choices allow regions of large $h$--fluxes, where $g_s$ is small while the scale $R$ determining the radii of the internal sphere and of the AdS spacetime is large. All in all, these solutions would thus appear a reliable starting point, but problems lurk around the corner and for one matter we recently ran across~\cite{gubmitr}, where the authors had reported on a violation of the Breitenlohner--Freedman bound~\cite{BF} for one of these cases, the $AdS_3 \times S^7$ solution of eq.~\eqref{ads3sol}, which would also resonate with some recent literature~\cite{weak_gravity}. These and other related issues are currently under investigation~\cite{basile}.

\section{Constrained superfields and a four--dimensional toy model}\label{sec:springoffs}

The last issue that we would like to address briefly is the emergence of interesting spring--offs of these and other investigations of broken supersymmetry. The subject proper targets non--linear realizations, albeit in the simpler context of four--dimensional models, and has a life of its own that started long ago, but nonetheless it connects to the preceding sections in an interesting fashion.

The one step forward that opened up a host of new possibilities was the recovery of the Volkov--Akulov model~\cite{VA}, which realizes the ${\cal N}=1 \to {\cal N}=0$ breaking in a non--linear regime, via a chiral superfield $\Phi$ subject to the \emph{quadratic} constraint~\cite{rocek,nonlinear,ks}
\beq
\Phi^2 \,=\,0 \ . \label{quadratic_constraint_WZ}
\eeq
This constraint replaces the complex scalar by a fermion bilinear, so that
\beq
\Phi \ = \frac{\psi\,\psi}{2\,F} \ + \ \sqrt{2}\,\theta^\alpha\,\psi_\alpha \ + \theta^2\,F \ .
\eeq
The Volkov--Akulov model then emerges from the Wess--Zumino Lagrangian with a Polonyi term, in the presence of the constraint~\eqref{quadratic_constraint_WZ}, up to a complicated field redefinition that was spelled out in~\cite{kt}.

Constrained superfields can also provide very instructive playgrounds in applications of supergravity to Cosmology, since for one matter they reduce to a minimum the fields involved in model building. This endeavour started with the example of the Starobinsky model~\cite{starobinsky}, whose realization~\cite{adfs} obtains coupling supergravity to an ordinary Wess--Zumino multiplet and a constrained one. Many interesting models then followed this work and~\cite{fkl}, which also include different realizations of the minimal coupling of supergravity to the Volkov--Akulov model~\cite{follow_ups}. These developments are reviewed in~\cite{nonlinear_reviews}, and the upshot is that, in all cases, one can apply the standard formula for the supergravity potential of~\cite{cfgvp},
\beq
{\cal V} \ = \ e^{\cal G} \left[ \left({\cal G}^{-1}\right)^{i \bar{j}} \, {\cal G}_i\,{\cal G}_{\bar{j}}  \ - \ 3\right]
\, \quad {\rm with} \qquad {\cal G} \ = \ {\cal K} \ + \ \log \left|{\cal W}\right|^2 \ , \eeq
while enforcing the constraint \eqref{quadratic_constraint_WZ} only at the end.

A second major step along this line was attained in the 1990's. It is the recovery of the supersymmetric Born--Infeld model, which realizes the ${\cal N}=2 \to {\cal N}=1$ breaking~\cite{bg}, via a \emph{quadratic} constraint linking to one another a chiral and a vector multiplet according to
\beq
{\cal C}_2: \ W^\alpha\, W_\alpha \ - \ 2\, \Phi \left( \frac{1}{4} \ \overline{D}^{\,2}\, \overline{\Phi} \ - \ m \right) \ = \ 0 \ . \label{2_1_constraint}
\eeq

The crucial difference between this case and the preceding one is that now, starting from ${\cal N}=2$, one ends up with ${\cal N}=1$ supersymmetry. There was some discussion as to whether this would be possible at all, and this development owes to the original considerations in~\cite{hughes_polchinski} and to the explicit linear models of~\cite{apt} and~\cite{cgfp}. In retrospect, the end result can be regarded as a toy model for a standard $D$ brane, whose insertion in the vacuum halves its original amount of supersymmetry. The supersymmetric Born--Infeld theory, with a scale determined by the real ``magnetic'' charge $m$, then emerges from the Polonyi term, since the constraint eliminates all kinetic terms in
\beq
{\cal L} \ = \ Re\left[ \frac{1}{2}\ \int d^2 \theta \, W^\alpha\,W_\alpha + (m \,-\,i\,e_c) \int d^2 \theta \,\Phi \right] \ + \ \int\,d^4 \theta \ {\overline \Phi} \, \Phi \ .
\eeq

These results also afford a natural generalization to the case of a number of Abelian multiplets, whereby the constraint of eq.~\eqref{2_1_constraint} becomes~\cite{fps}
\beq
d_{ABC} \left[ W^{\alpha\,B}\, W_\alpha^C \ - \ 2\, \Phi^B \left( \frac{1}{4} \ \overline{D}^{\,2}\, \overline{\Phi}^C \ - \ m^C \right)\right] \ = 0 \ ,
\eeq
which might provide some useful information on the long--sought non--Abelian Born--Infeld theory associated to $D$-brane stacks~\cite{tseytlin_review}. It would be interesting to characterize its microscopic origin and its role in String Theory.

Let us conclude with a more recent result, described in~\cite{dfs}, which concerns the non--linear ${\cal N}=2 \to {\cal N}=0$ breaking in a vector multiplet, and is therefore a simple four--dimensional toy model of brane supersymmetry breaking. The relevant constraint is now cubic, and reads
\beq
 \Phi\, W^\alpha\, W_\alpha \ - \ \Phi^2
\left( \frac{1}{4} \ \overline{D}^{\,2}\, \overline{\Phi}
\ - \ m \right) \ = \ 0 \ \longrightarrow \  \Phi^3 \ =
 \ 0 \, , \ \Phi^2\, W_\alpha \ = \ 0 \ .
 \eeq
One can show that, aside from a singular corner where the ${\cal N}=2 \to {\cal N}=1$ breaking is recovered, this \emph{cubic} constraint eliminates the complex scalar of the Wess-Zumino multiplet, leaving the gauge field and the two fermions of the original multiplets, which are the two goldstini of the system. Notice that, in contrast with the ${\cal N}=1 \to {\cal N}=0$ case of~\cite{bg}, here a Born--Infeld structure of the remaining low--energy interactions is not implied. Other recent works dealing with the ${\cal N}=2 \to {\cal N}=0$ case, also from the vantage point of Volkov--Akulov model, can be found in~\cite{recent20}.

We have addressed these developments from the vantage point of brane supersymmetry breaking, but the four--dimensional counterpart of the tadpole potentials in eq.~\eqref{string_frame_bsb} and \eqref{einstein_frame_bsb} is also the uplift potential of the KKLT construction~\cite{KKLT}. The work of~\cite{kvw}, where combinations of different constrained ${\cal N}=1$ multiplets are used to characterize the four--dimensional content of the associated branes, is also close in spirit to this simple example.

\vskip 14pt
\noindent {\large \bf Acknowledgements}

\noindent J.M. is grateful to Scuola Normale and A.S. is grateful to CPhT--\'Ecole Polytechnique, APC--U. Paris VII and CERN-TH for the kind hospitality while the work reviewed here was in progress. A.S. was supported in part by Scuola Normale Superiore and by INFN (IS CSN4-GSS-PI). We are also grateful
to C.~Angelantonj, I.~Basile, E.~Dudas, S.~Ferrara, A.~Gruppuso, N.~Kitazawa, M.~Lattanzi, N.~Mandolesi, P.~Natoli and G.~Pradisi for extensive discussions and/or collaborations on related issues and/or comments on the manuscript.
\vskip 14pt
\setcounter{equation}{0}

\end{document}